\documentclass[a4paper,fleqn,usenatbib]{mnras}
\usepackage{newtxtext}
\usepackage[varvw]{newtxmath}
\usepackage[T1]{fontenc}
\usepackage{ae,aecompl}
\usepackage[dvipdfmx]{graphicx}	
\usepackage{amsmath}	
\usepackage{here}
\usepackage{bm}
\usepackage[dvipsnames]{xcolor}

\newcommand{\RED}[1]{{ #1}}

\title[Fragmentation with Lagrangian hydrodynamics]{Testing the effect of resolution on gravitational fragmentation with Lagrangian 
hydrodynamic schemes
} 
\author[Y. Yamamoto, T. Okamoto \& T. Saitoh]{
Yasuyoshi Yamamoto$^{1}$, 
Takashi Okamoto$^{1}$\thanks{E-mail: okamoto@astro1.sci.hokudai.ac.jp}, 
Takayuki R. Saitoh$^{2, 3}$
\\
$^{1}$Faculty of Science, Hokkaido University, N10 W8, Kitaku, Sapporo, 060-0810, Japan\\
$^{2}$Department of planetology, Graduate school of science, Kobe
University, 101, Rokkodai-cho, Nada-ku, Kobe, Hyogo, 657-8501, Japan\\
$^{3}$Earth-Life Science Institute, Tokyo Institute of Technology, 2-12-1 Ookayama, Meguro-ku, Tokyo, 152-8551, Japan\\
}

\date{Accepted XXX. Received YYY; in original form ZZZ}

\pubyear{2020}

\begin{document}
\label{firstpage}
\pagerange{\pageref{firstpage}--\pageref{lastpage}}
\maketitle

\begin{abstract} To study the resolution required for simulating
gravitational fragmentation with newly developed Lagrangian hydrodynamic
schemes, Meshless Finite Volume method (MFV) and Meshless Finite Mass method
(MFM), we have performed a number of simulations of the Jeans test and
compared the results with both the expected analytic solution and results
from the more standard Lagrangian approach: Smoothed Particle Hydrodynamics
(SPH). We find that the different schemes converge to the analytic solution
when the diameter of a fluid element is smaller than a quarter of the Jeans
wavelength, $\lambda_\mathrm{J}$. Among the three schemes, SPH/MFV shows the
fastest/slowest convergence to the analytic solution. \RED{Unlike the
well-known behaviour of Eulerian schemes,} none of the Lagrangian schemes
investigated displays artificial fragmentation when the perturbation
wavelength, $\lambda$, is shorter than $\lambda_\mathrm{J}$, even at low
numerical resolution. \RED{For larger wavelengths ($\lambda >
\lambda_\mathrm{J}$) the growth of the perturbation is delayed when it is not
well resolved. Furthermore, with poor resolution, the fragmentation seen with
the MFV scheme proceeds very differently compared to the converged solution.
All these results suggest that, when unresolved, the ratio of the magnitude
of hydrodynamic force to that of self-gravity at the sub-resolution scale is
the largest/smallest in MFV/SPH, the reasons for which we discussed in
detail.} These tests are repeated to investigate the effect of kernels of
higher-order than the fiducial cubic spline. Our results indicate that the
standard deviation of the kernel is a more appropriate definition of the
`size' of a fluid element than its compact support radius.
\end{abstract}




\section{Introduction}

In the formation of astronomical objects, such as stars and galaxies, self-gravity plays a central role. The non-linearity of these formation processes is one of the main reasons numerical simulations remain an essential part of astrophysical research at a variety of scales. Knowing the numerical requirements to follow gravitational collapse and fragmentation correctly by any specific numerical scheme is, therefore, of utmost importance.

\citet{truelove1997} showed, by using an \RED{Eulerian} adaptive mesh
refinement (AMR) finite difference method, that the cell size $d$ must
satisfy the Jeans condition, $d < \lambda_\mathrm{J}/4$, where
$\lambda_\mathrm{J} = (\pi c_\mathrm{s}^2/G \rho)^{1/2}$ is the local Jeans
length, $c_\mathrm{s}$ is the local sound speed, $G$ is the gravitational
constant, and $\rho$ is the local gas density. They claim that artificial fragmentation can occur when this condition is not satisfied in simulations using the finite difference method.
\RED{Throughout this paper, we use the term `artificial fragmentation' as in
\citet{truelove1997}. Namely, artificial fragmentation is the fragmentation
of stable perturbations, which occurs when the Jeans length is not resolved or is only marginally resolved. This fragmentation vanishes when one simulates the same problem with a higher resolution that resolves the Jeans length. }

\citet{bate1997} derive the corresponding Jeans 
condition for Smoothed Particle Hydrodynamics (SPH; \citealt{lucy1977}; \citealt{gingold1977}) 
\RED{; SPH is inherently a Lagrangian scheme, while \citet{truelove1997} use an Eulerian scheme to derive 
the Jeans condition.}
They argue that, if the gravitational softening length 
is set to be comparable to the particle smoothing length, the minimum mass resolved with SPH is
$M_\mathrm{MIN} \simeq N_\mathrm{NEIB} m$, where $N_\mathrm{NEIB}$ is the
number of neighbouring particles used for the SPH calculation and $m$ is the
mass of an SPH particle. They show that, in order to \RED{obtain converged results with SPH}, the local Jeans mass,
\begin{equation} M_\mathrm{J} \equiv \frac{4 \pi (\lambda_\mathrm{J}/2)^3
\rho}{3} = \frac{\pi^{5/2} c_\mathrm{s}^3}{6 G^{3/2}\rho^{1/2}},
\end{equation} 
must always be resolved \citep{bate1997,bate2002}. 

\citet{hubber2006} further investigate the issue of spurious fragmentation in SPH. They simulate a test
problem, called the Jeans test, and show that failing to resolve the local
Jeans length only delays physical fragmentation, rather than inducing
artificial collapse. SPH is thus a robust scheme in this context.
\citet{okamoto2003} and \citet{agertz2007}, however, find that SPH has a
fundamental difficulty in dealing with the mixing of two distinctive phases
at contact discontinuities. Their findings motivated further improvements of the SPH technique by a number of different authors \citep[e.g.][]{price08, wadsley2008, read2010, saitoh2013, kawata2013}.

Recently, new classes of Lagrangian hydrodynamic schemes\footnote{Strictly
speaking, none of these schemes are genuine Lagrangian methods. The term
`Lagrangian' here only means that each fluid element moves with the local
fluid velocity.} have been developed such as a moving 
mesh methods \citep{springel2010, TESS, RICH} 
and mesh-free methods \citep{gaburov2011, hopkins2015, gandalf}.
Despite appearing somewhat akin to the Lagrangian-based SPH, it is dangerous to simply extrapolate our knowledge of SPH convergence properties to these new schemes.

Therefore, we aim to test the resolution requirements to correctly model 
gravitational fragmentation for these new mesh-free methods by utilising the Jeans test 
of \citet{hubber2006}. The schemes we test in this paper are the Meshless 
Finite-Mass (MFM) and Meshless Finite-Volume (MFV) methods, both implemented in 
the publicly available code {\scriptsize GIZMO} \citep{hopkins2015}.
We also include the original  ``traditional", density-based, SPH approach (TSPH) 
to compare to legacy Lagrangian methods. 
We also use a modern algorithmic implementation of SPH as a comparative 
benchmark, specifically the pressure-based SPH (PSPH) scheme 
\citep[PSPH,][]{saitoh2013, hopkins2013} originally proposed by
\citet{ritchie2001}. In PSPH, the smoothed variable is pressure (or internal energy)
instead of using the traditional density-based approach, with density instead 
obtained from the smoothed pressure via the equation of state; doing this solves 
the well-known surface tension problem in SPH \citep{saitoh2013}. 
All the methods (i.e. MFM, MFV,  TSPH and PSPH) are implemented in {\scriptsize GIZMO} 
and hence we can focus on  the differences due to the hydrodynamic schemes. 

Modern SPH schemes also tend to employ higher-order kernels with several hundred
neighbours \citep[e.g.][]{read2010, dehnen2012} instead of the standard
cubic spline kernel with $\sim 32$ neighbours to reduce the so-called E0 
error\footnote{The zero-th order error.} while avoiding the pairing instability.
Therefore, we also carry out the Jeans
test simulations with these higher-order kernels. Throughout this paper we
employ the adaptive gravitational softening proposed by \citet{price2007},
which manifestly conserves momentum and energy, allowing for comparable
resolution in self-gravitational and hydrodynamic calculations. Changing the
kernel function for the hydrodynamic calculations thus changes the shape of
gravitational softening, so we will also explore the effective resolution that is
needed to \RED{obtain converged results} for these higher-order kernels.
 
The structure of this paper is as follows: In Section~\ref{simulation}, we
briefly describe the Jeans test and our simulation setup. We present our
results in Section~\ref{results} and we discuss and summarise our main conclusions in
Section~\ref{conclusions}.

\section{Simulations} \label{simulation}

In this section, we briefly describe the Jeans test itself, our adopted kernel functions, and how initial conditions are created to mimic those of \citet{hubber2006}.

\subsection{The Jeans test} \label{jeans test}

In order to investigate the effect of numerical resolution on gravitational fragmentation, we use the Jeans test of \citet{hubber2006} that
has a \RED{linearised} analytic solution \citep{jeans1929}. In this test, we
consider a static infinite medium, with uniform density $\rho_{0}$ and uniform and constant isothermal sound speed $c_\mathrm{s}$. We then impose a perturbation so that $\rho_{0}\rightarrow\rho_{0}+\rho_{1}$ and
$\bm{v}_{0}=\bm{0}\rightarrow\bm{v}_{0}+\bm{v}_{1}=\bm{v}_{1}$, such that the \RED{linearised} continuity, Euler, and Poisson equations are: 
\begin{align} 
	&\frac{\partial\rho_{1}}{\partial t} = 
	-\rho_{0}\bm{\nabla}\cdot\bm{v}_{1} \label{eq:1}, \\ 
	&\frac{\partial\bm{v}_{1}}{\partial 
	t}=-\frac{c_{s}^{2}\bm{\nabla}\rho_{1}}{\rho_{0}}-\bm{\nabla}\phi_{1} 
	\label{eq:2}, \\ 
	&\bm{\nabla}^{2}\phi_{1}=4\pi G\rho_{1} 
	\label{eq:3},
\end{align}
where $\phi_{1}$ is the gravitational potential due to the perturbed density.
By combining these three equations, we obtain the equation of motion for the 
density perturbation,  
\begin{equation} \frac{\partial^2 \rho_1}{\partial t^2} - c_\mathrm{s}^2
\bm{\nabla}^2 \rho_1 - 4 \pi G \rho_0 \rho_1 = 0. \end{equation}%
By substituting a one-dimensional plane wave solution, 
$\rho_{1}(\bm{r},t)=A\rho_{0}e^{i(k x\pm\omega t)}$, 
we obtain the dispersion relation: 
\begin{equation}
\omega_{k}^{2} = c_{\mathrm{s}}^{2} k^{2} - 4\pi G\rho_{0},
\label{eq:dispersion}
\end{equation}
resulting in a critical wave-number of
$k_\mathrm{J} = (4 \pi G \rho_0)^{0.5}/c_\mathrm{s}$. 

The corresponding critical wavelength, the Jeans wavelength, is
$\lambda_\mathrm{J} \equiv (\pi c_{\mathrm{s}}^{2}/G\rho_{0})^{1/2}$. 
Equation~(\ref{eq:dispersion}) can be written by using the 
Jeans wavelength as 
\begin{equation}
\omega_{\lambda}^{2} =4 \pi^{2}c_{s}^{2}\left(\frac{1}{\lambda^{2}}-\frac{1}{\lambda_\mathrm{J}^{2}}\right).
\label{eq:4}
\end{equation}

As in \citet{hubber2006}, we superimpose two plane waves of equal amplitude
and wavelength, travelling in opposite directions, to set up an initially stationary plane wave perturbation,
\begin{align}
\rho_{1}(\bm{r},t)&=\frac{A\rho_{0}}{2}\left\{e^{i(kx-\omega_{k}t)}+e^{i(kx+\omega_{k}t)}\right\}, 
\label{eq:5}
\\
\bm{v}_1(\bm{r}, t) &= \frac{A\omega}{2 k}\left\{e^{i(kx - \omega_k t)} - e^{i(kx+\omega_k t)}\right\} \hat{\bm{e}}_x, 
\label{eq:vel1}
\end{align}
where $\hat{\bm{e}}_x$ is the unit vector pointing in the $x$-direction. We also define the resulting density fluctuation as
\begin{equation}
	\delta(\bm{r}, t) \equiv \frac{\rho_1(\bm{r}, t)}{\rho_0}  
\end{equation}
for convenience. 
For a short wavelength perturbation $(\lambda<\lambda_{J})$, 
$\omega_\lambda^{2} $ 
is positive (Eq.~(\ref{eq:4})), and therefore the perturbation oscillates with 
the period:
\begin{equation}
T_{\lambda} = \left(\frac{\pi}{G\rho_{0}}\right)^{1/2}\frac{\lambda}{
	(\lambda_{J}^{2} - \lambda^{2})^{1/2}}.
\label{eq:oscilation}
\end{equation}
For a long wavelength perturbation $(\lambda>\lambda_{J})$,
$\omega_\lambda^{2}$ is negative, and the perturbation instead grows.
\citet{hubber2006} characterise the growth timescale as the time for the
perturbed density on the plane $x=0$ to grow from $A\rho_{0}$ to
$\cosh(1)A\rho_{0}$. This timescale is given as 
\begin{equation}
T'_{\lambda}\;=\;\left(\frac{1}{4\pi G\rho_{0}}\right)^{1/2}\frac{\lambda}{(\lambda^{2}-\lambda_\mathrm{J}^{2})^{1/2}}.
\label{eq:contraction}
\end{equation}
These analytic timescales can then be compared these to  those measured in simulations with different numerical resolutions and hydrodynamic schemes. \RED{We note that above timescales are derived from the set of linearised equations, whereas the simulation code solves the non-linear equations governing the fluid evolution.}

\subsection{Initial conditions}

By inserting $t = 0$ into Eqs.~(\ref{eq:5}) and (\ref{eq:vel1}) and 
taking the real parts, we obtain the initial states for  
both short and long wavelength perturbations: 
\begin{align}
\rho(\bm{r},0) &= \rho_{0}\left\{1\;+\;A \cos\left(\frac{2\pi\mathit{x}}{\lambda}\right)\right\}, 
\label{eq:perturbed_density}\\
\bm{v}(\bm{r},0) &= 0. 
\label{eq:9}
\end{align}

To set up an initial condition, we first realize a uniform density
distribution with a glass-like fluid element configuration. We randomly
distribute $N_\mathrm{tot}$ fluid elements within a unit cube, and allow the system to evolve without self-gravity using periodic boundary conditions. This
reduces the Poisson density fluctuations, and produces an approximately
uniform, but non-crystalline, density distribution. We wait until the relative density errors become smaller than 1\% before any perturbation is applied. Note that, for the equal mass fluid elements, the definition of density is identical among all the schemes investigated in this study. The mean kernel size of a fluid element is given by
\begin{equation}
	\bar{h}=\left(\frac{3 N_\mathrm{NEIB}}{4\pi N_{\mathrm{tot}}}\right)^\frac{1}{3}. 
	\label{eq:smoothing}
\end{equation}
For a fixed $N_\mathrm{NEIB}$, we can change the resolution by changing $N_\mathrm{tot}$ since the lengths defining the cubic simulation volume are always unity.

Once the system is sufficiently relaxed, we add a perturbation to this uniform density field. We impose a one-dimensional 
sinusoidal density perturbation by adjusting the unperturbed $x$-coordinate,
$x_{i}$, of each fluid element, $i$, to a perturbed one $x_{i}'$.
We use the following relation to transform the
coordinates as in \citet{hubber2006},
\begin{equation}
x_{i}'+\frac{A\lambda}{2\pi}\sin\left(\frac{2\pi x_{i}'}{\lambda}\right)=x_{i}, 
\label{eq:11}
\end{equation}
where the fractional amplitude, $A$, is set to 0.1, and $\lambda$ is parameter used 
to vary the size of the imposed perturbation. 
We solve this equation iteratively. \RED{We note that this
imposed density perturbation is quite large from the perspective of the
linear theory. Nevertheless, \citet{hubber2006} showed that converged solutions are well characterised by the timescales obtained by the linearised equations.}

The density perturbation calculated from this perturbed distribution of the fluid 
elements via Eq.~(\ref{eq:sph_density}) or (\ref{eq:mf_density}) is 
smaller than that given by Eq.~(\ref{eq:perturbed_density}) 
at low resolution as we will demonstrate later (Section~\ref{accuracy}. 
However, we first employ these initial conditions to carry out the same test as \citet{hubber2006} for the mesh-free methods.

\subsection{Kernel functions}
\begin{figure}
	\centering
  \includegraphics[width=\columnwidth]{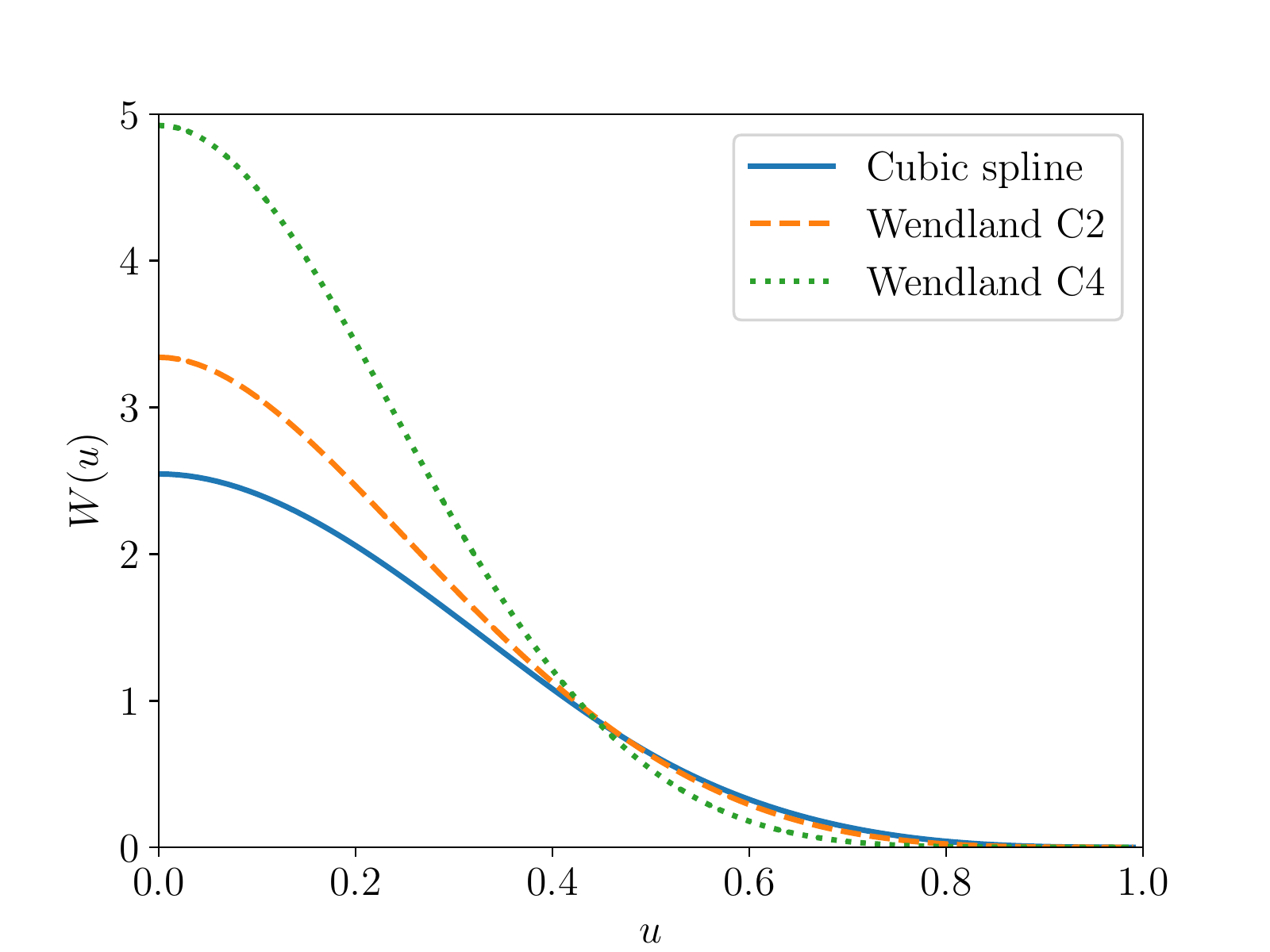}
	\caption{
		Shapes of the kernel functions used in this work for $h_i = 1$. 
		\RED{The horizontal axis, $u$, is defined as $u = r/h_i$ where 
		$r$ is the distance from the fluid element $i$.}
		The cubic spline, Wendland C2, and Wendland C4 kernels are indicated by 
		the solid, dashed, and dotted lines, respectively. 
	}
	\label{fig:kernel_functions}
\end{figure}

In our simulations we employ 3 different kernel functions. The first is the standard cubic spline kernel 
\citep{monaghan1985}: 
\begin{equation}
W(u,h_{i}) = \frac{8}{\pi h_i^3}\begin{cases}
1+6 u^{2}(u-1) & \mathrm{for}~0\le u < \frac{1}{2}, \\
2(1-u)^{3} & \mathrm{for}~\frac{1}{2}\le u <1, \\
0 & \mathrm{otherwise},
\label{eq:cubic}
\end{cases}
\end{equation}
where $h_i$ is the kernel size (or smoothing length) of a fluid element, and $u$
is defined as $u \equiv |\bm{x}-\bm{x_{i}}|/h_{i}$, where $\bm{x}_i$ is the position of
the fluid element.  
Modern SPH prescriptions often employ higher order Wendland functions \citep{wendland1995} as kernels to avoid the well-known paring instability when using a large number of neighbours, which is preferable as it reduces low-order errors in the SPH method \citep{dehnen2012}. In this paper, we employ the Wendland C2 and 
C4 kernels in addition to the standard cubic spline kernel. The functional form of the three-dimensional Wendland C2 kernel is:
\begin{equation}
W(u,h_{i}) = \frac{21}{2\pi h^{3}_{i}}\begin{cases}
(1-u)^{4}(1+4u) & \mathrm{for}~0\le u<1, \\
0 & \mathrm{otherwise},  
\label{eq:wendC2}
\end{cases}
\end{equation}
and that of the Wendland C4 kernel is: 
\begin{equation}
W(u,h_{i}) = \frac{495}{32\pi h^{3}_{i}}\begin{cases}
(1-u)^{6}(1+6u+\frac{35}{3}u^{2}) & \mathrm{for}~0\le u<1, \\
0 & \mathrm{otherwise}. 
\label{eq:wendC4}
\end{cases}
\end{equation}
These three kernel functions used in this study are shown in 
Fig.~\ref{fig:kernel_functions}, which clearly shows that 
the higher-order Wendland kernels are more centrally concentrated 
than the cubic spline. Since higher-order kernels assign more weight 
to closer fluid elements, they generally require higher values of 
$N_\mathrm{NEIB}$ to correctly reproduce the actual fluid density \citep{dehnen2012}. 

\RED{In {\scriptsize GIZMO}, the kernel size, $h_i$, is determined such that the smoothed number density of fluid elements, $n(\bm{x}_i)$, is constant
(see also \citealt{springel2002}), given by:
\begin{equation}
	n(\bm{x}_i) \equiv \sum_j W(|\bm{x}_j - \bm{x}_i|/h_i, h_i) = \frac{N_\mathrm{NEIB}}{(3/4)\pi h_i^3}.
\end{equation}
which is solved iteratively for $h_i$.
The density of an SPH particle is then: 
\begin{equation}
	\rho_i^\mathrm{SPH} = \sum_j m_j W(|\bm{x}_j - \bm{x}_i|/h_i, h_i), 
	\label{eq:sph_density}
\end{equation}
where $m_j$ is the mass of the $j$-th SPH particle. 
In MFM and MFV, the density of a fluid element is defined instead as 
\begin{equation}
	\rho_i^\mathrm{MFM/MFV} = m_i n(\bm{x}_i),  
	\label{eq:mf_density}
\end{equation}
because $n(\bm{x}_i)^{-1}$ is the volume of the fluid element in these two schemes. When fluid elements all have the same mass, as in our initial conditions, these two definitions above are identical. 
}

\subsection{The definition of resolution}

As in \citet{hubber2006}, we define the minimum resolvable mass as 
$M_{\mathrm{MIN}}=N_\mathrm{NEIB} m$. 
The Jeans condition states that the Jeans mass $M_\mathrm{J}$ must 
exceed $M_\mathrm{MIN}$, that is, 
\begin{equation}
M_{\mathrm{MIN}}=N_{\mathrm{NEIB}}m \leq M_\mathrm{J}=\frac{4}{3}\pi\left(\frac{\lambda_\mathrm{J}}{2}\right)^{3}\rho=\frac{\pi^{5/2}c_\mathrm{s}^{3}}{6G^{3/2}\rho^{1/2}}. 
\label{eq:12}
\end{equation}
This condition can also be read as effectively setting a density threshold of: 
\begin{equation}
\rho\leq\left(\frac{\pi}{6 N_{\mathrm{NEIB}}m}\right)^{2}\left(\frac{\pi c_{s}^{2}}{G}\right)^{3}.
\label{eq:13}
\end{equation}
By using Eq.(\ref{eq:smoothing}) as the kernel size, we derive the following condition,
\begin{equation}
\lambda_\mathrm{J} \geq 2 \bar{h},
\label{eq:14}
\end{equation}
namely, the Jeans wavelength should exceed the diameter of a fluid element. 
This Jeans Condition is identical to that used in \citet{hubber2006}.

We define the resolution, $\mathcal{R}$, as a ratio of the mean diameter of a 
fluid element, $\bar{d}=2\bar{h}$, to the wavelength of 
the perturbation, 
$\lambda=n^{-1}_{\lambda}$, where $n_{\lambda}$ is the integer number of wavelengths 
that fit within a side of the simulation box (of length $= 1$), i.e.
\begin{equation}
\mathcal{R}=\frac{\bar{d}}{\lambda}=n_{\lambda}2\bar{h}=n_{\lambda}\left(\frac{6{N}_{\mathrm{NEIB}}}{\pi N_{\mathrm{tot}}}\right)^{1/3}.
\end{equation}
A smaller value of $\mathcal{R}$ corresponds to better resolution. 
For $\lambda = \lambda_\mathrm{J}$, 
the Jeans condition (Eq.~(\ref{eq:14})) reads: 
\begin{equation}
\mathcal{R}\leq 1. 
\end{equation}
%
%
%
For the tests presented in the following sections, we fix the perturbation wavelength at each resolution and alter the ratio $\lambda/\lambda_\mathrm{J}$ by changing the isothermal sound speed, $c_\mathrm{s}$.

\section{Results} \label{results}
\begin{figure}
	\centering
	\includegraphics[width=\columnwidth]{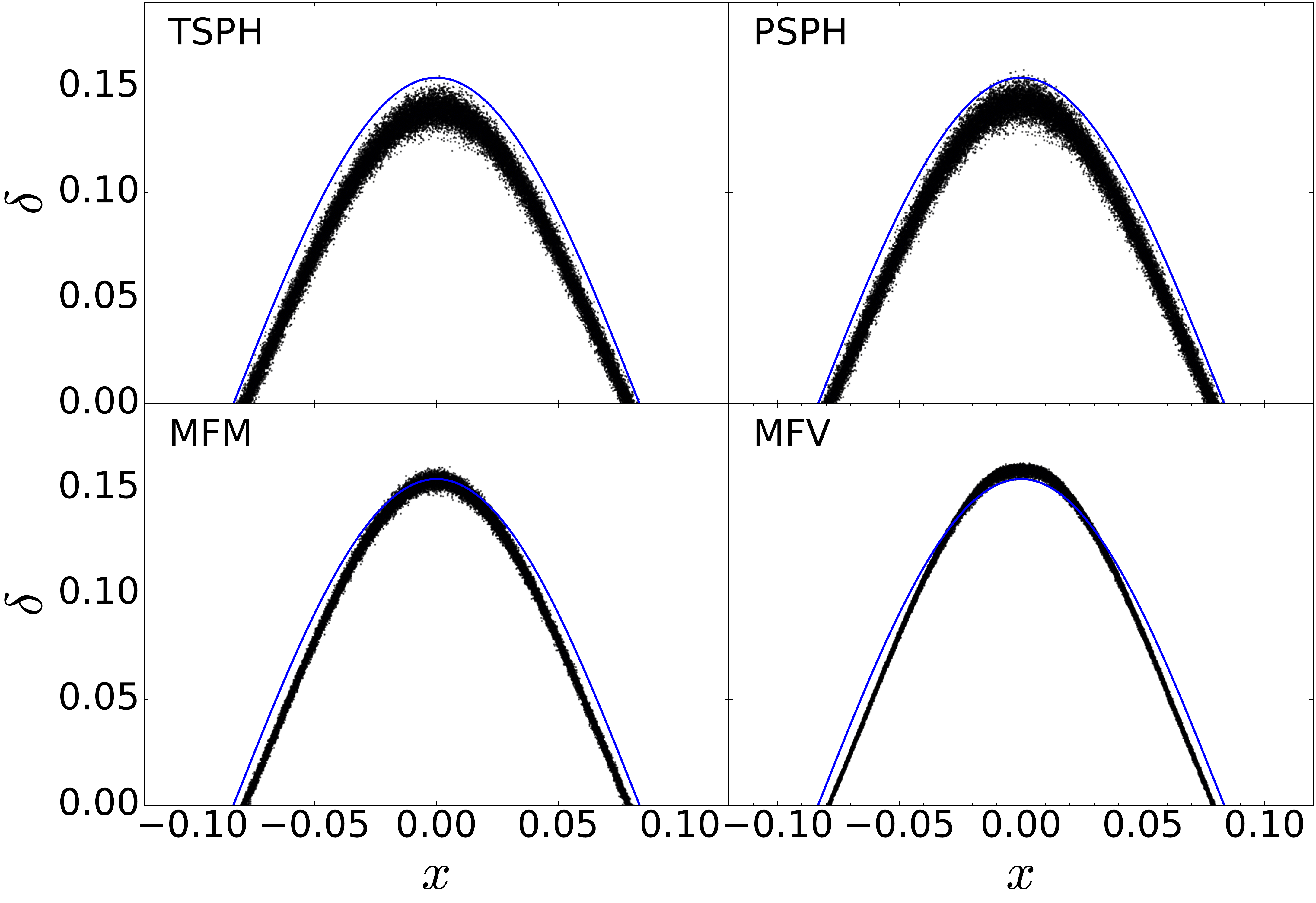}
	\caption{
		The density fluctuation, $\delta$, as a function of $x$-position of each fluid element when the maximum
		density fluctuation, $\max\{\delta_i(t)\}$, reaches $\cosh(1)$
		times the initial maximum density perturbation,
		$\max\{\delta_i(t=0)\}$. The
		resolution of each simulation is $\mathcal{R}=0.25$ and the
		wavelength of a perturbation is $\lambda/\lambda_\mathrm{J}=1.2$. The
		upper-left, upper-right, lower-left, and lower-right panels present
		the results of TSPH, PSPH, MFM, and MFV, respectively. The blue solid
		line shows the density fluctuation obtained by multiplying the initial
		density fluctuation by $\cosh(1)$.
	}
	\label{fig:Comparison_growth}
\end{figure}

In this section we report the results of the simulations outlined in Section\,\ref{simulation}. We measure the oscillation period for small scale perturbations ($\lambda < \lambda_\mathrm{J}$) and the characteristic timescale for the maximum density fluctuation to increase by $\cosh(1)$ for the larger scale perturbations that trigger collapse ($\lambda < \lambda_\mathrm{J}$), and compare both to the analytic solutions of Eq.~(\ref{eq:oscilation}) and  Eq.~(\ref{eq:contraction}).

In Fig.~\ref{fig:Comparison_growth}, we show the density fluctuations of the fluid elements against their $x$-positions when the maximum density fluctuation,
$\delta^\mathrm{max}(t) \equiv \max\{\delta_i(t)\}$, in the simulation box reaches $\cosh(1)$ times the initial maximum density fluctuation, $\delta^\mathrm{max}(t=0)$, for the four different hydrodynamical schemes for the specific case of $\mathcal{R} = 0.25$ (well-resolved) and $\lambda/\lambda_\mathrm{J} = 1.2$ (collapsing).

We find that, due to the large noise in SPH, the maximum density
in TSPH and PSPH does not represent the typical density at the location of the peak.
Because of this, we underestimate the characteristic timescale for SPH 
when we simply use the maximum density. 
On the other hand, the density distribution in MFV is very tight, indicating 
that MFV is the least diffusive method of the four. 
Note that the initial density fields are identical in all simulations. 

To circumvent this problem, we evaluate the mean density at the $x$-position 
of a fluid element, $i$, by using fluid elements within 
$0.1 h_i$, i.e. $|x_j - x_i| < 0.1 h_i$.
Doing so makes the density distribution a single-valued function of $x$ 
and we avoid the bias caused by the noisy density field present in the SPH methods. 
We call this density $\tilde{\rho}$ to discriminate it from  the density of a 
fluid element. 
We also define $\tilde{\delta}$ as the density fluctuation computed from $\tilde{\rho}$. 
This mean density, $\tilde{\rho}$, is used to define all the growth timescales quoted 
in this paper.

\subsection{Comparisons between different Lagrangian methods}\label{comparison}

\begin{figure}
	\centering
  \includegraphics[width=\columnwidth]{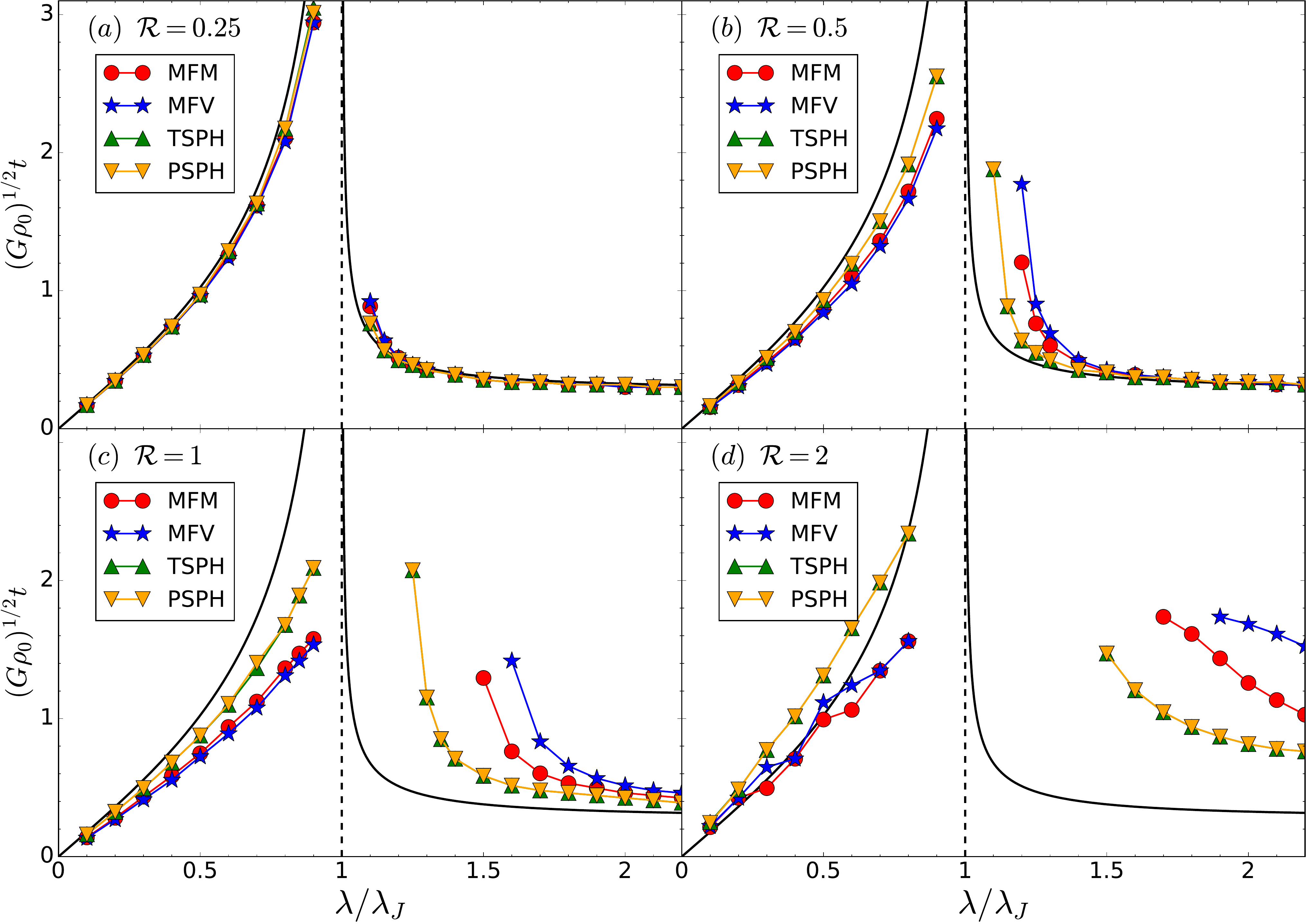}
	\caption{
		Oscillation periods for $\lambda < \lambda_\mathrm{J}$ and 
		growth time scales for $\lambda > \lambda_\mathrm{J}$ as 
		functions of the perturbation wavelength normalised by the Jeans 
		wavelength. The circles, stars, upward triangles and downward 
		triangles, respectively, indicate results from the MFM, MFV, TSPH, 
		and PSPH methods. The analytic solutions, Eq.~(\ref{eq:oscilation}) and 
		Eq.~(\ref{eq:contraction}), are represented by the solid lines.  
		In panels (a), (b), (c), and (d), we present results with the 
		resolution, $\mathcal{R} = 0.25$, 0.5, 1, and 2, respectively. 
	}
	\label{fig:Comparison1}
\end{figure}

In Fig.~\ref{fig:Comparison1}, we show the results of our Jeans test 
simulations. 
In all simulations presented here, we employ the cubic spline kernel and 
$N_\mathrm{NEIB} = 50$. 
Firstly, we confirm the earlier results by \citet{hubber2006}, that
is, (i) perturbations which should oscillate ($\lambda < \lambda_\mathrm{J}$)
always oscillate and do not display artificial fragmentation even when the
Jeans mass is not resolved ($\mathcal{R} = 2$), (ii) poor resolution
stabilizes unstable perturbations ($\lambda > \lambda_\mathrm{J}$) near
$\lambda \sim \lambda_\mathrm{J}$ and the wavelength of the perturbations that
fail to grow becomes longer with poorer resolution (i.e. larger
$\mathcal{R}$), and (iii) poorer resolution makes the growth timescale
longer for a given wavelength.

Comparing the results from the four different schemes, TSPH and PSPH 
are indistinguishable at all resolutions. This is always true in the simulations 
presented in this paper, and hence we hereafter simply refer to them collectively as SPH. 
In the highest resolution simulations ($\mathcal{R} = 0.25$), 
the characteristic timescales obtained with different schemes nicely converges to the analytic timescales and there is little difference between them. 

In lowering the resolution the numerical results deviate from the analytic
estimates.
Except for the oscillating perturbations in the lowest resolution simulations
($\lambda < \lambda_\mathrm{J}$ and $\mathcal{R} = 2$), the results with SPH
are closer to the analytic estimates than MFM and MFV. The oscillation periods with MFV are shorter than those with SPH for $\lambda <
\lambda_\mathrm{J}$ and the growth timescales with MFV are longer than those
with SPH, indicating that the ratio of the magnitude of the hydrodynamic force to that of 
self-gravitational force in MFV is larger than in SPH. The results with MFM consistently lie between SPH and MFV, and always closer to MFV.

For the growing perturbations, the growth timescale described by
Eq.~(\ref{eq:contraction}) is a downwardly convex function. \RED{When the 
perturbation is not resolved ($\mathcal{R} = 2$), the growth timescale with
MFV does not display this characteristic form, while the curve obtained with SPH
is still downwardly convex at this resolution. MFM is somewhat ambiguous at this resolution, while generally exhibiting the desired downwardly convex behaviour.} 
This may suggest some unphysical behaviour of MFV when the Jeans condition is violated. We shall revisit this issue in Section~\ref{accuracy}.

\subsection{Comparisons with different kernel functions}
\begin{figure*}
	\centering
  \includegraphics[width=\linewidth]{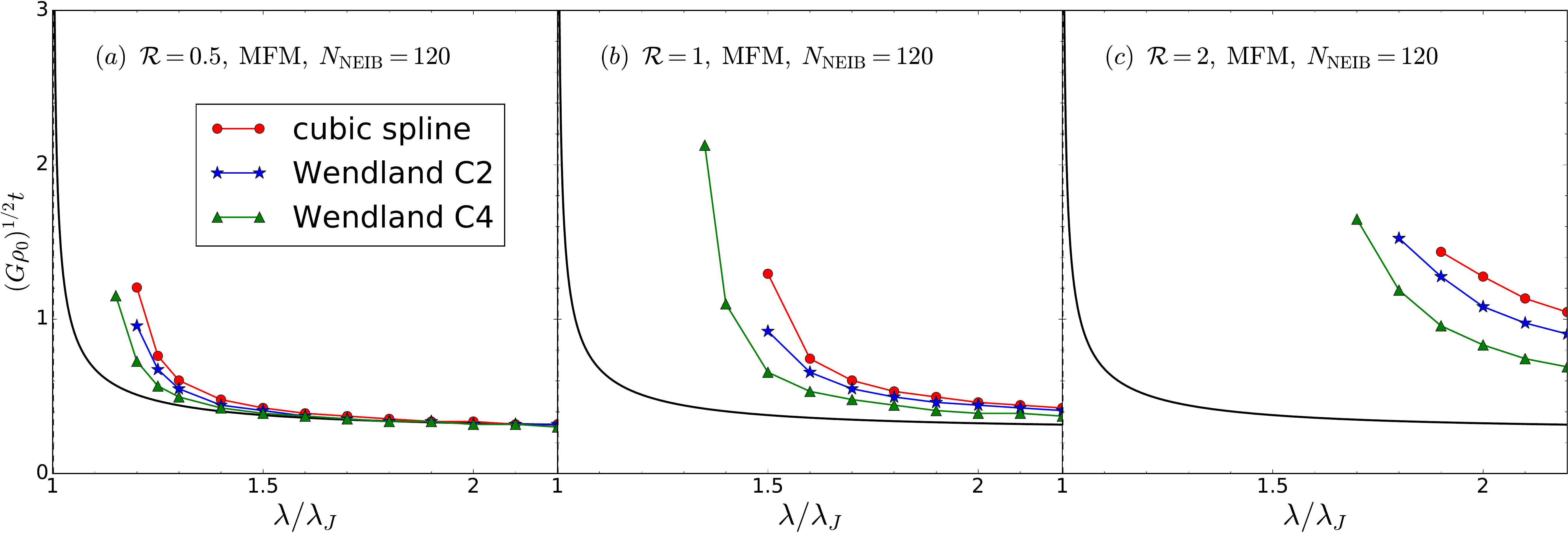}
	\caption{
		The growth timescales with different kernel functions and the MFM method. 
		The circles, stars, and triangles indicate the results with 
		the cubic spline, Wendland C2, and Wendland C4 kernels, 
		respectively. 
		Form left to right, the resolutions adopted are 
		$\mathcal{R} = 0.5$, 1.0, and 2.0. 
	}
	\label{fig:Comparison2}
\end{figure*}
\begin{figure}
	\centering
  \includegraphics[width=\columnwidth]{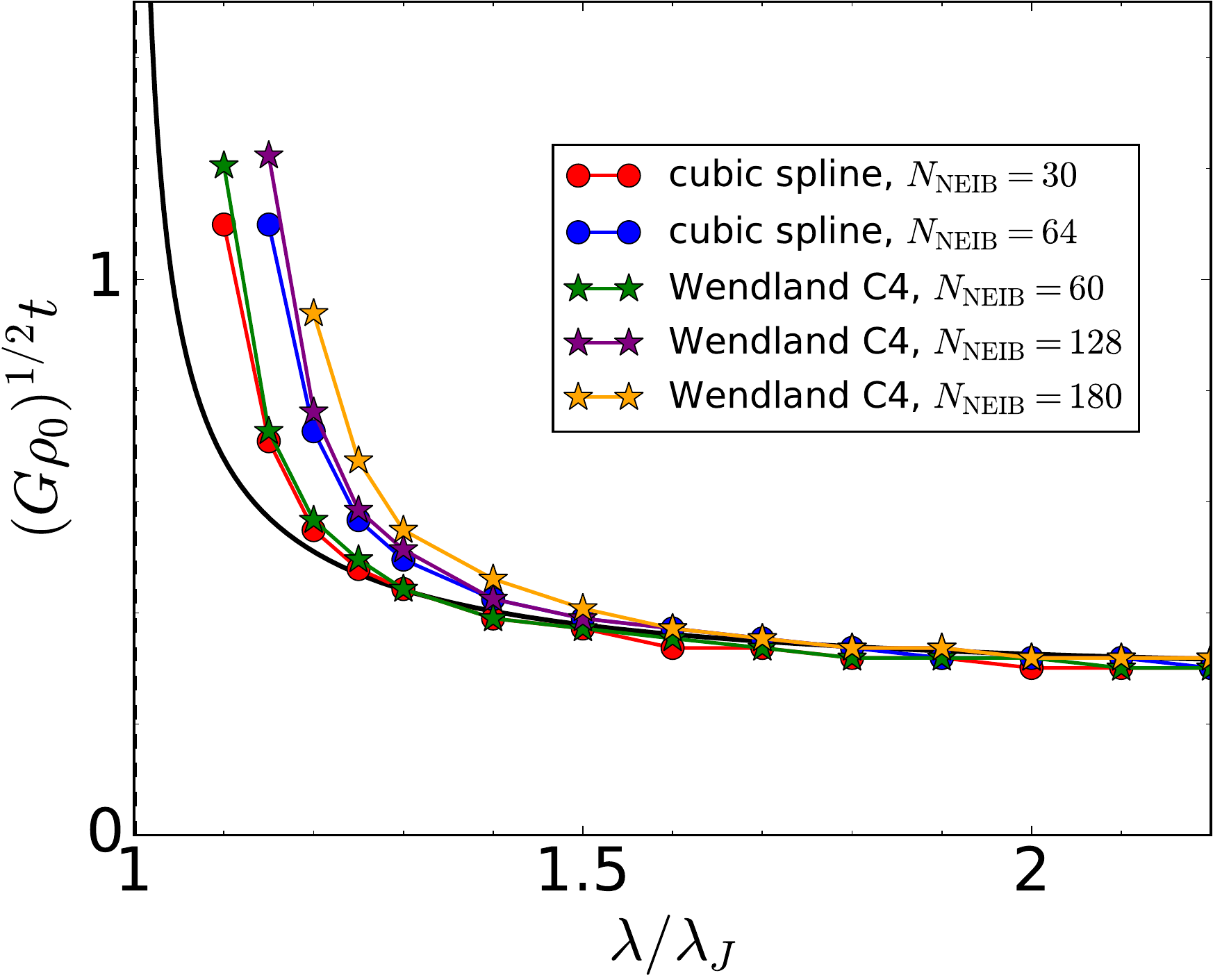}
	\caption{
		The growth timescales with several combinations of the kernel functions 
		and the number of neighbour particles. 
		The circles and stars represent the results by the cubic spline and Wendland C4 kernels, respectively. 
		For the cubic spline kernel, the line connecting the symbols 
		closer to the analytic estimate indicates the results with 
		$N_\mathrm{NEIB} = 30$ and the other line with $N_\mathrm{NEIB} = 64$. 
		For the Wendland C4 kernel, the lines indicate 
		$N_\mathrm{NEIB} = 60$, 128, and 180 from the closet to the analytic
		estimate to the farthest, respectively. 
	}
	\label{fig:Comparison3}
\end{figure}
In this subsection, we again carry out the Jeans test, but now investigating the effect of changing the kernel function. 
We employ MFM as the hydrodynamic scheme in this subsection. Firstly, we fix $N_\mathrm{NEIB}$ to 120 to focus on the effect of different
kernel shapes. This number of neighbours is typical for the Wendland C4
kernel, and may otherwise cause the paring instability if used with the cubic spline kernel in SPH \citep[e.g.][]{dehnen2012}. MFM and MFV should be free from this
problem and we can safely use the cubic spline kernel with such a large number of
neighbours \citep{hopkins2015}.

In Fig.\ref{fig:Comparison2}, we show the growth timescales with the cubic
spline, Wendland C2, and Wendland C4 kernels by changing the resolution from
$\mathcal{R} = 0.5$ to 2.0. The results with MFM are similar to those shown
in Fig.~\ref{fig:Comparison1}. 
For a given resolution, $\mathcal{R}$, the
timescales obtained with the higher-order kernels are closer to the analytic
estimate than their lower-order counterparts. As is evident from
Fig.~\ref{fig:kernel_functions}, higher-order kernels are more centrally
concentrated than lower-order ones and therefore allocate more weight to closer
fluid elements. Also, the gravitational force is less softened by the higher-order kernels for a given kernel size.

Our results indicate that defining resolution as the ratio of 
the mean diameter or kernel size of a fluid element to the local Jeans wavelength 
is not appropriate when several kernel functions are in use. 
It is therefore convenient to introduce a definition of resolution that is independent of the kernel function. 

We compare the growth timescales with the cubic spline and Wendland C4 kernels
by changing $N_\mathrm{NEIB}$ in Fig.~\ref{fig:Comparison3}, in which we use 
$N_\mathrm{NEIB} = 30$ and $64$ for the cubic spline kernel and $N_\mathrm{NEIB} = 
60$, 128 and 180 for the Wendland C4 kernel.

We find that the results using the cubic spline kernel with $N_\mathrm{NEIB} = 
30$ and 64 are almost identical to those by the Wendland C4 kernel with 
$N_\mathrm{NEIB} = 60$ and 128, respectively. 
This indicates that the resolutions $\mathcal{R} \simeq 0.32$ and 0.41 for the 
cubic spline kernel roughly correspond to $\mathcal{R} \simeq 0.4$ and 0.51 
for the Wendland C4 kernel, respectively. 
In other words, the effective kernel size of the cubic spline kernel is $\sim
2^{1/3} = 1.26$ times as large as that of the Wendland C4 kernel for a 
given kernel size. 

\citet{dehnen2012} defined the effective hydrodynamic resolution as the 
kernel standard deviation, $\sigma$: 
\begin{equation}
	\sigma^2 = \nu^{-1} \int \mathrm{d}^\nu x \ \bm{x}^2 W(|\bm{x}|/h, h),
	\label{eq:sd}
\end{equation}
where $\nu$ is the number of spatial dimensions. 
The effective smoothing length $h'$ is then defined as 
\begin{equation}
	h' = 2 \sigma. 
\end{equation} 
The ratio of the effective smoothing length to the kernel size, $h'/h$, is
$\simeq 0.5477$ and $\simeq 0.4529$ for the cubic spline and Wendland C4
kernels, respectively \citep{dehnen2012}. 
The effective smoothing length of the cubic spline
kernel is thus $\sim 1.21$ times as large as that of the Wendland C4 kernel
for a given kernel size. This scaling is similar to that indicated by
Fig.~\ref{fig:Comparison3}. Since the effective smoothing length is about
half the kernel size, we can define the effective resolution, $\mathcal{R}'$,
in a kernel-independent manner as 
\begin{equation} 
	\mathcal{R}' = \frac{4 h'}{\lambda}. 
\end{equation}
The results shown in Fig.~\ref{fig:Comparison1} can be applied to other kernels 
by replacing $\mathcal{R}$ with $\mathcal{R}' = 1.095 \mathcal{R}$. 

\subsection{Effects of the amplitude of initial perturbations}
\label{accuracy}

\begin{figure}
	\centering
	\includegraphics[width=\columnwidth]{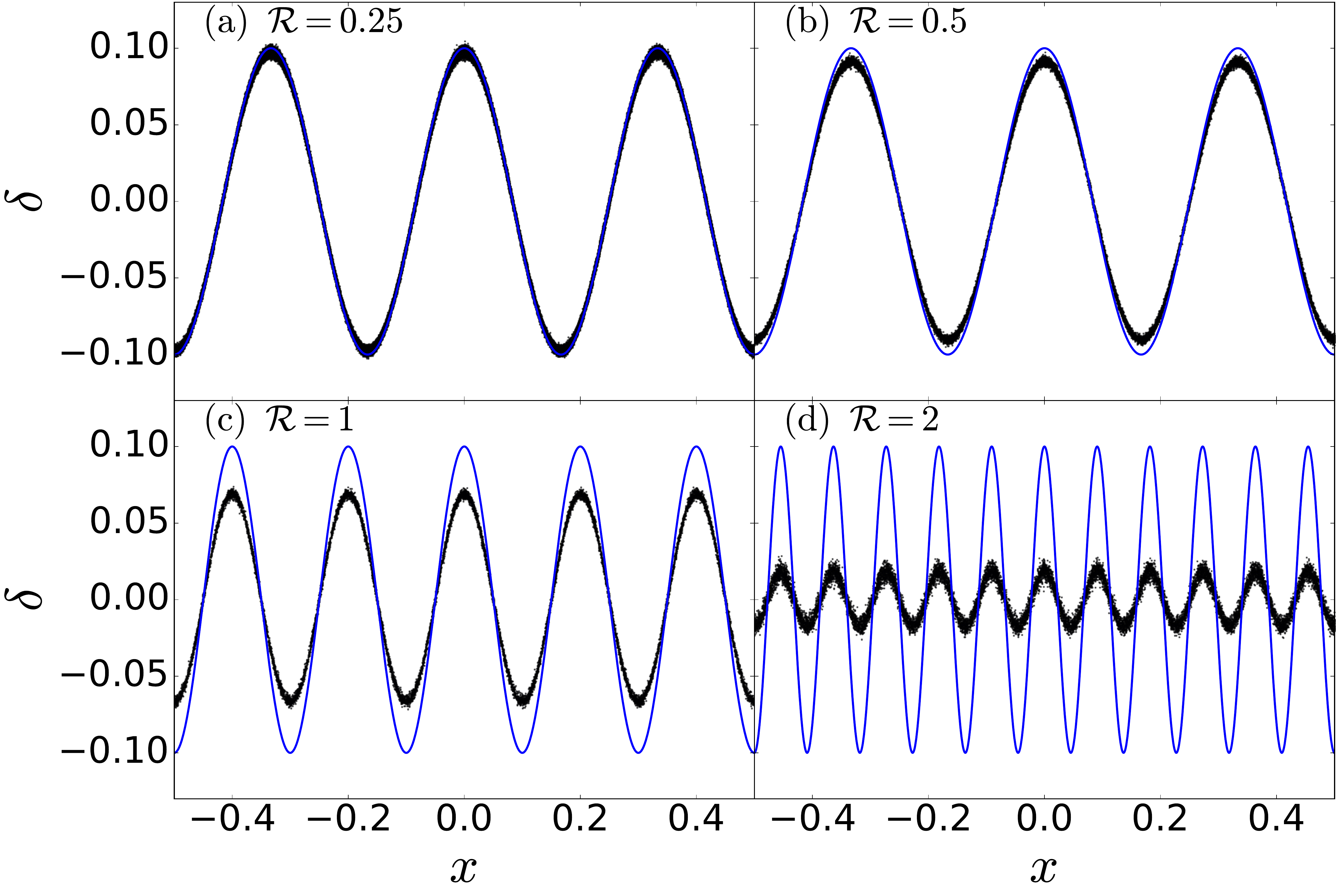}
	\caption{The initial density fluctuation at each resolution. 
	The density fluctuation of each fluid element is shown against its 
	$x$-position by black points. 
	The blue solid line represents the target density fluctuation given by 
	Eq.~(\ref{eq:perturbed_density}) with $A = 0.1$.
	The panels (a), (b), (c), and (d) show the initial conditions for 
	 resolutions of $\mathcal{R} = 0.25$, 0.5, 1, and 2, with the 
	cubic spline kernel, respectively. 
	}
	\label{fig:Comparison_ics}
\end{figure}

So far, we have used the same coordinate transformation as
\citet{hubber2006} to generate the initial perturbations (Eq.~(\ref{eq:11})).
The initial perturbations added in this way, however, depend on the 
resolution. 
In Fig.~\ref{fig:Comparison_ics}, we show the density fluctuation of each 
fluid element with the cubic spline kernel and the target density given by 
Eq.~(\ref{eq:perturbed_density}) with $A = 0.1$.  
We find that the higher the resolution, the better the target initial 
condition is reproduced. 
At the lowest resolution, $\mathcal{R} = 2$, the amplitude is only one-third
of the target amplitude due to the kernel size being much larger than the
wavelength of the perturbations. This small amplitude should not affect our
measure of the timescales because we define the characteristic growth
timescale as the time at which the maximum density fluctuation reaches
$\cosh(1)$ times the initial maximum density fluctuation. Nevertheless, it is
interesting to investigate the response when the imposed initial
density perturbation more accurately matches the targeted amplitude.

\begin{figure}
	\centering
	\includegraphics[width=\columnwidth]{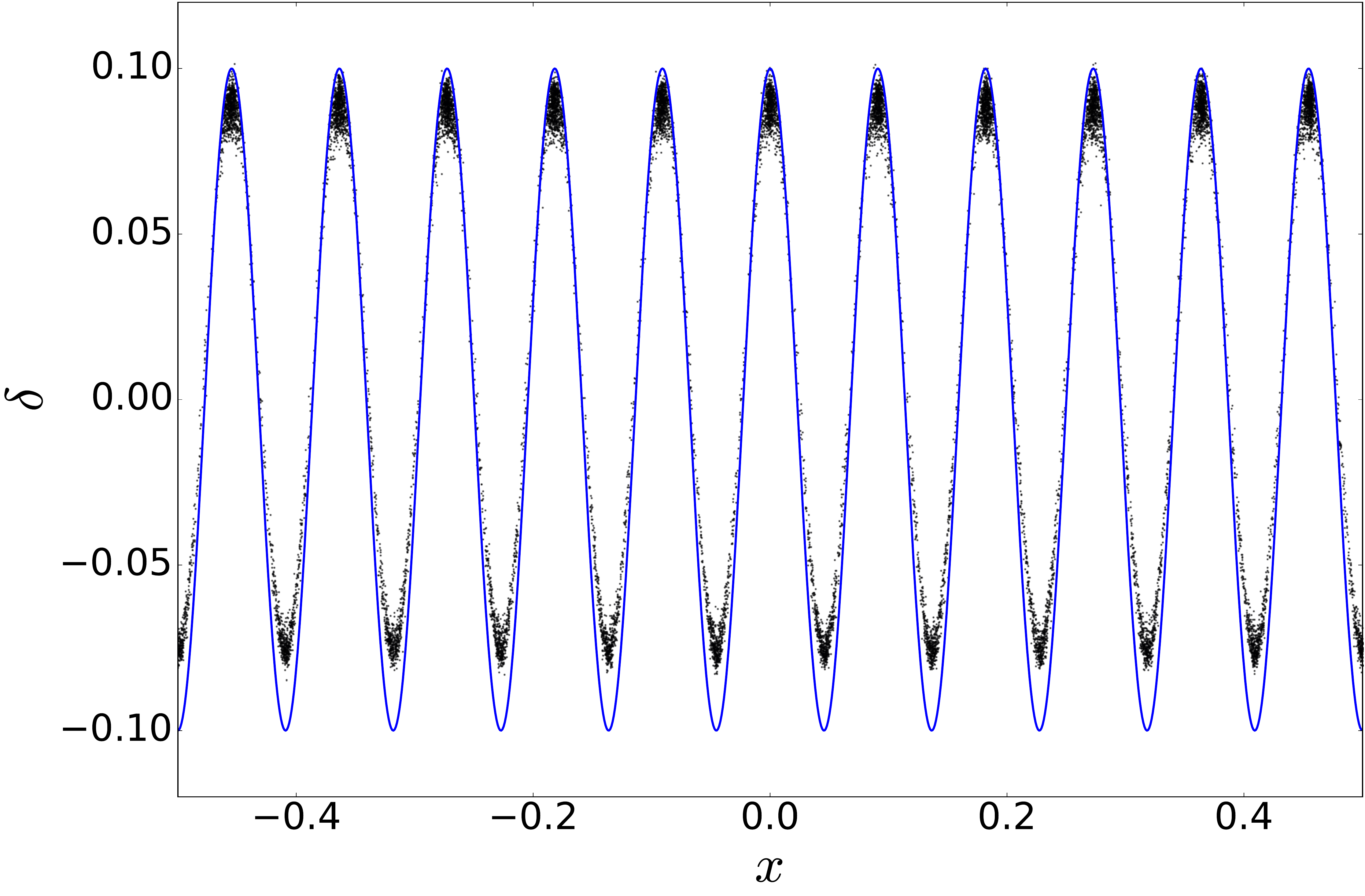}
	\caption{
		The same as Fig.~\ref{fig:Comparison_ics} but showing the revised 
		initial condition for $\mathcal{R} = 2$. 
	}
	\label{fig:Revised_ics}
\end{figure}
To generate an appropriate density distribution, we evolve the system by
adjusting the pressure of each fluid element without self-gravity. When a
fluid element has higher/lower density than the target density at its
position, $x$, we increase/decrease its pressure according to the following
equation:
\begin{equation}
	P_i = \bar{P} \left(1 + \Delta \frac{\rho_i - \rho_\mathrm{t}(x_i)}{\lambda}\right), 
	\label{eq:pressure}
\end{equation}
where $\rho_\mathrm{t}(x_i)$ is the target density at the position of 
the $i$-th particle given by Eq.~(\ref{eq:11}) with $A = 0.1$, $\bar{P}$ is 
the reference pressure which we can arbitrarily determine, and $\Delta$ is a adjustable parameter.   
We start from the original initial condition and evolve the system by 
adjusting the pressure through Eq.~(\ref{eq:pressure}) with relatively large values $\Delta$. 
We then gradually decrease the value of $\Delta$ until  
the total relative variation, 
$\sum_i |\rho_i - \rho_\mathrm{t}(x_i)|/\rho_\mathrm{t}(x_i)$, 
cannot become smaller anymore. 

We show the density fluctuation obtained by this procedure in 
Fig.~\ref{fig:Revised_ics} for $\mathcal{R} = 2$.  
The density distribution is now much closer to the target density than 
the original initial condition obtained by the coordinate transformation of Eq. ({\ref{eq:11}}).

\begin{figure}
	\centering
  \includegraphics[width=\columnwidth]{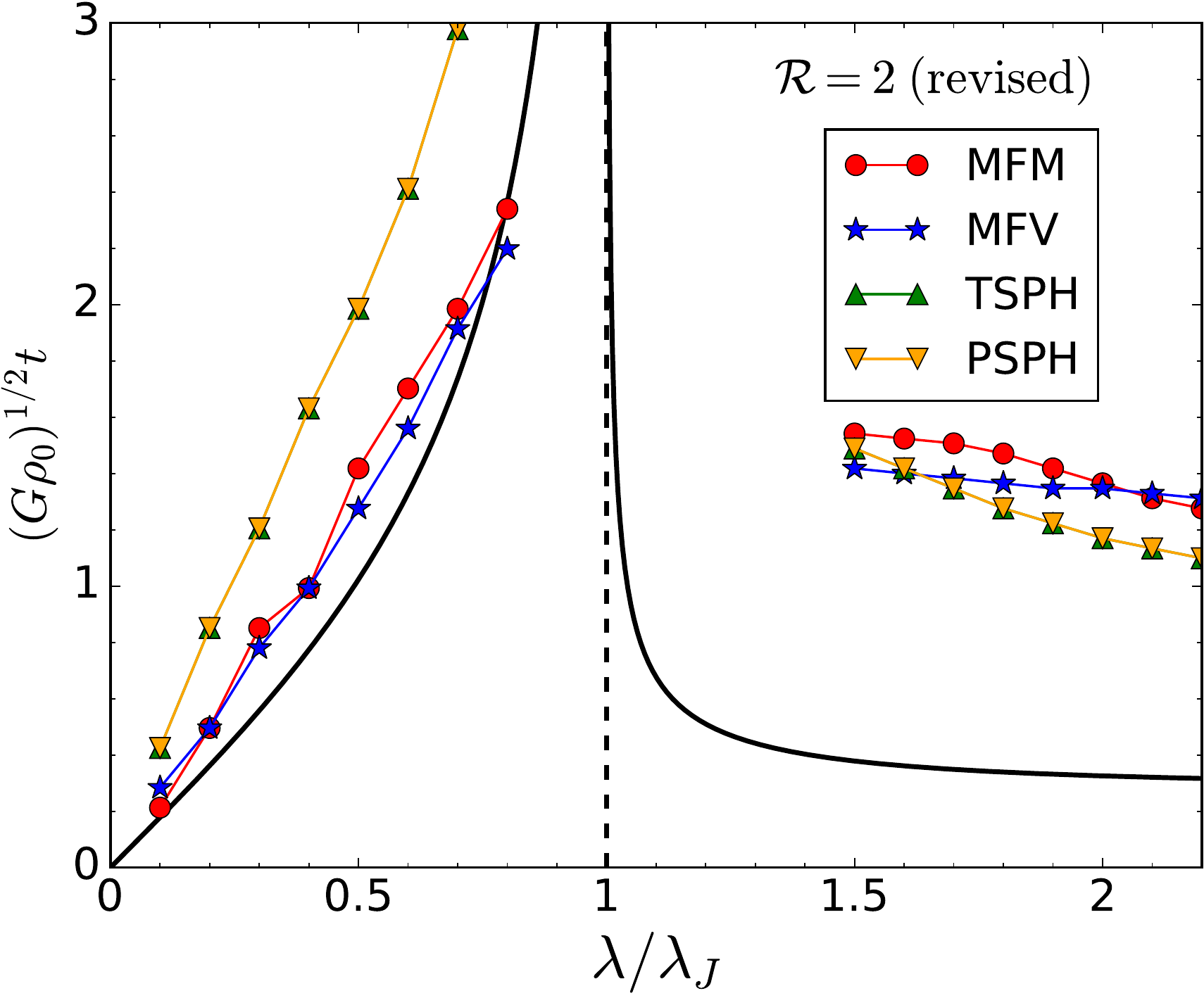}
	\caption{
		The same as Fig.~\ref{fig:Comparison1} but with the revised initial 
		conditions. 
		Only the results for $\mathcal{R} = 2$ are shown.
		}
	\label{fig:Jeans_test_revised}
\end{figure}

Using initial conditions generated in this manner, we again perform the Jeans 
test simulations and vary the hydrodynamical scheme. We show the results in Fig.~\ref{fig:Jeans_test_revised}. 
We find that the results noticeably change from the ones shown in 
Fig.~\ref{fig:Comparison1}. 

For the short wavelength perturbations ($\lambda<\lambda_{J}$), 
the oscillation periods become longer than those with the original initial 
conditions, while the slopes of the timescales as functions of wavelength 
are maintained.  
For the long wavelength perturbations $(\lambda_{J}<\lambda)$, 
the difference between the hydrodynamic methods becomes smaller than 
that shown in Fig.~\ref{fig:Comparison1}. 
The timescales obtained with MFV are almost independent of wavelength. 

\begin{figure}
	\centering
	\includegraphics[width=\columnwidth]{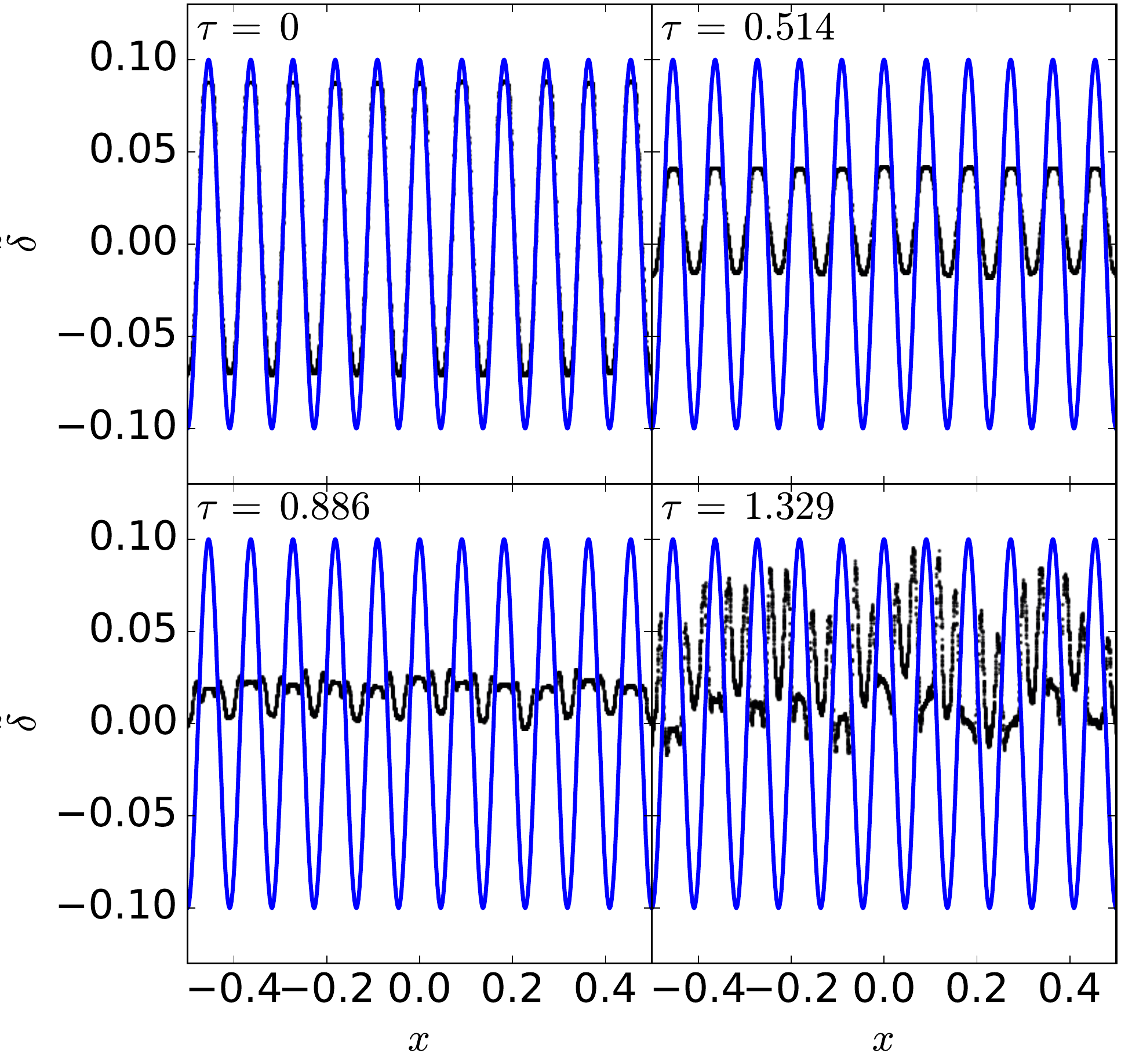}
	\caption{The time evolution of the density fluctuation profile 
	in a simulation with MFV. 
	We use the cubic spline kernel and a resolution of 
	$\mathcal{R}=2$, with an initial perturbation of $\lambda/\lambda_\mathrm{J} = 1.5$. We also show the initial density fluctuation given by 
	Eq.~(\ref{eq:perturbed_density}) with $A = 0.1$ as blue solid lines. 
	The time is expressed as $\tau=(G\rho_{0})^{1/2}t$. 
	}
	\label{fig:time_develop}
\end{figure}
\begin{figure*}
	\centering
	\includegraphics[width=14cm]{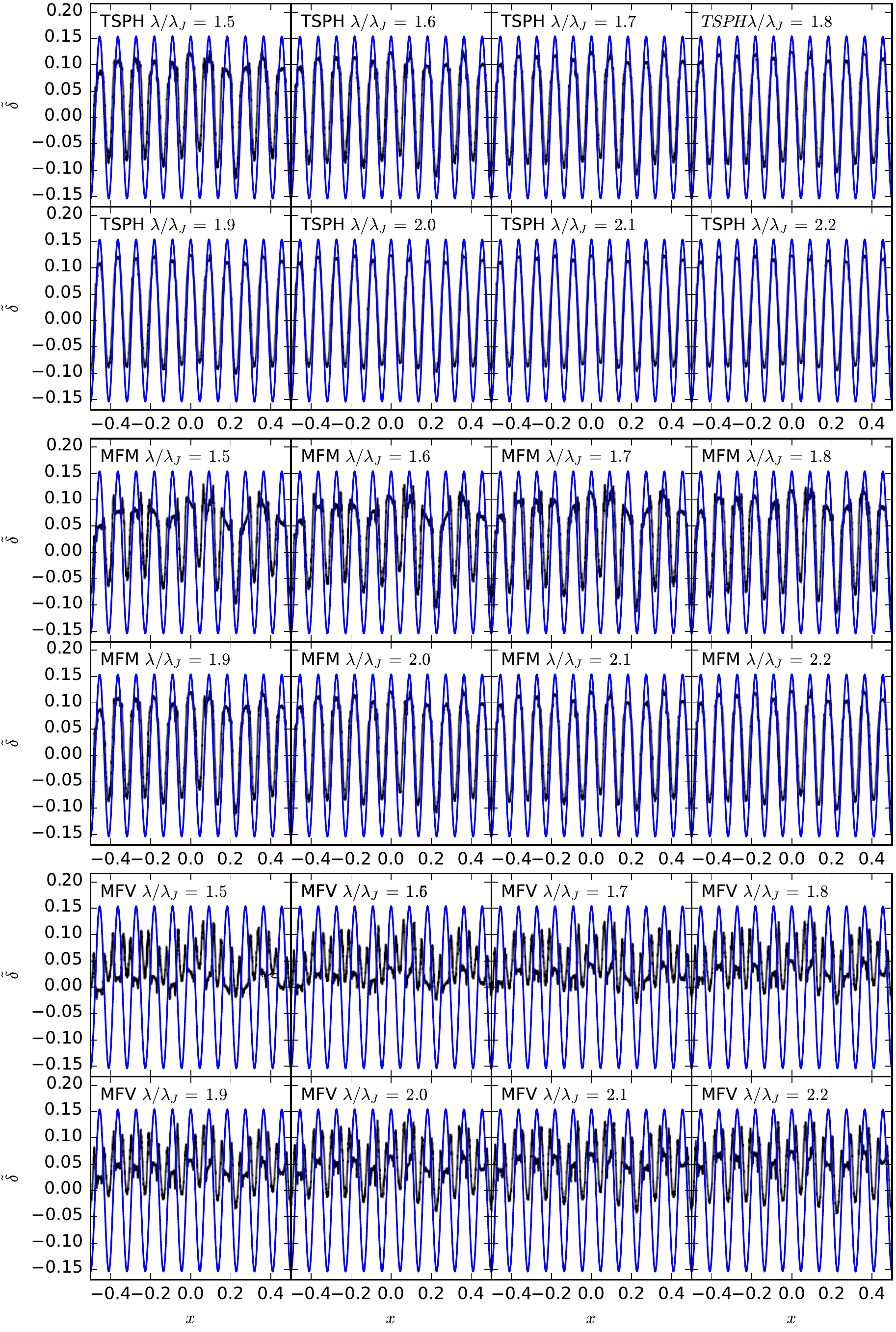}
	\caption{
		The density fluctuation profiles, $\tilde{\delta}(x)$, 
		when the maximum density fluctuation reaches $\cosh(1)$
		times the initial maximum density fluctuation 
		at the resolution of $\mathcal{R} = 2$ with the cubic spline kernel. 
		From top to bottom,
		the results are shown using the TSPH, MFM, and MFV methods, respectively. The blue solid line indicates the density field obtained by multiplying the density fluctuation in Eq.~(\ref{eq:perturbed_density}) by $\cosh(1)$.
	}
	\label{fig:final}
\end{figure*}
To understand this odd behaviour in the growth timescale when using MFV, we
show the explicit time evolution of $\tilde{\delta}(x)$ of the fluid elements with position, $x$, 
in Fig.~\ref{fig:time_develop}. 
The resolution is $\mathcal{R} = 2$ and the wavelength of the perturbation is
$\lambda/\lambda_\mathrm{J} = 1.5$.

We find that the perturbations initially do not grow but decay. They almost
disappear by $\tau = 0.886$, then shorter wavelength perturbations start to
grow (see $\tau = 1.329$). These density fluctuations are the ones that reach
$\cosh(1)$ times the initial maximum density fluctuation when we define the
characteristic timescale. 
\RED{We note that we do not observe this behaviour of initial decay and subsequent 
fragmentation into short-wavelength perturbations when we run simulations from 
the original initial conditions, in which the amplitude of the initial perturbation 
is much smaller than intended.  (see Fig.~\ref{fig:Comparison_ics}~(d)). 
When the amplitude of the unresolved perturbation with $\lambda > \lambda_J$ 
is too small as the original one and its wavelength is close to the Jeans 
length, the perturbation oscillates.}

In Fig.~\ref{fig:final}, we show the density fluctuation profiles, 
$\tilde{\delta}(x)$, at the final epoch, at which $\delta^\mathrm{max}(t) =
\cosh(1) \delta^\mathrm{max}(t=0)$), with TSPH, MFM, and MFV with changing perturbation wavelength in the range of $\lambda/\lambda_\mathrm{J} = 1.5$ to
2.2. We find that results with TSPH are robust, 
\RED{at least the sinusoidal shapes are 
always maintained}, even when we strongly violate the
Jeans condition; only a slight signature of wave deformation is seen
when the perturbation wavelength is close to the Jeans length.
With MFV, perturbations always strongly deform and the density peaks are not 
located at those in the initial conditions,  
even when the perturbation wavelength is 2.2 times the Jeans length. 
MFM again shows an intermediate behaviour between SPH and MFV, though 
closer to that of SPH. 

\begin{figure*}
	\centering
	\includegraphics[width=16cm]{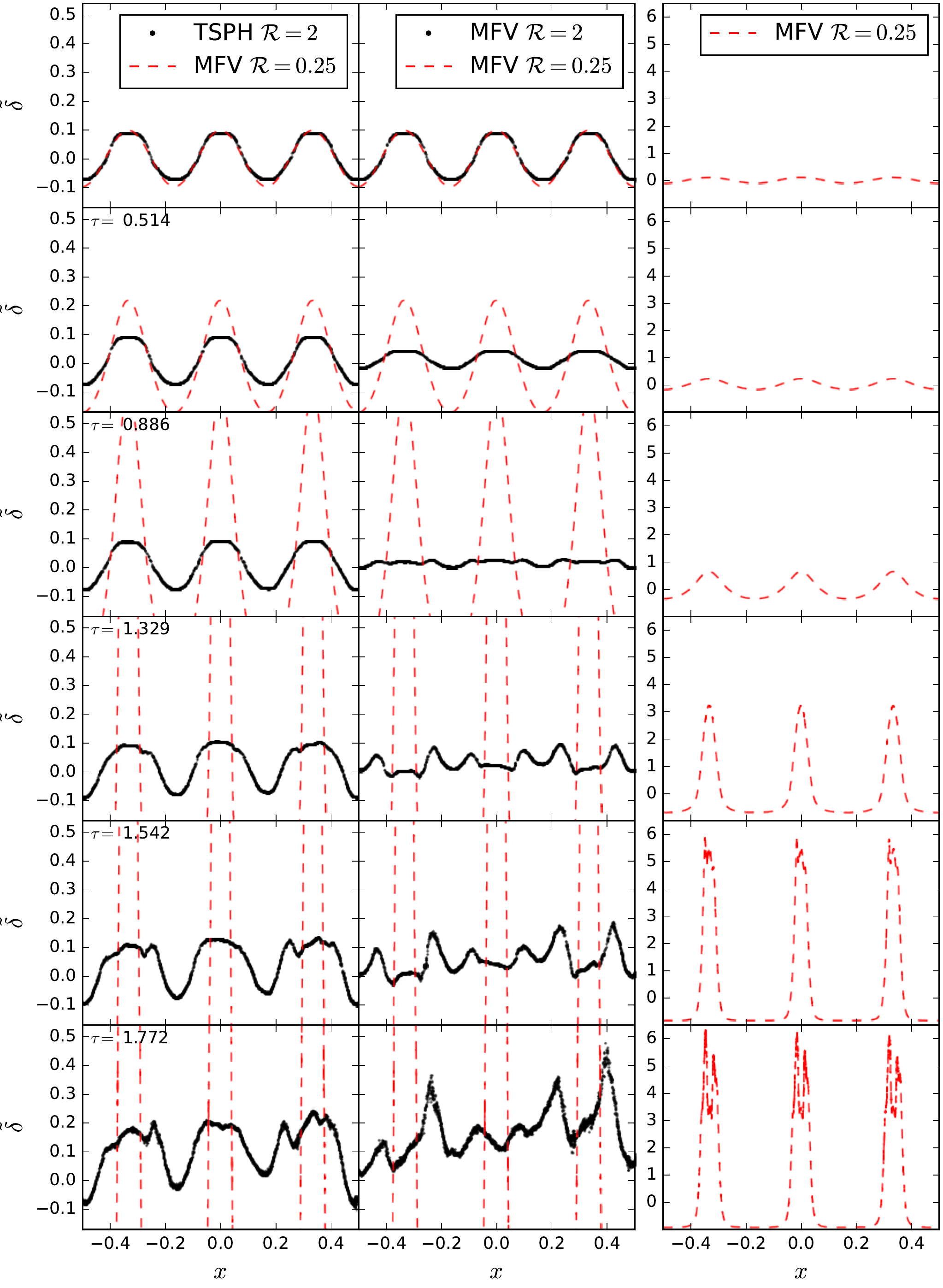}
	\caption{The time evolution of the density fluctuation profiles
	, $\tilde{\delta}(x)$, in SPH and MFV are shown in the left 
	and middle columns, respectively, as black dots. The resolution for these simulations is $\mathcal{R} = 2$
	and the perturbation wavelength is $\lambda/\lambda_\mathrm{J} = 1.5$. We
	also show the resolved simulation ($\mathcal{R} = 0.25$) with MFV as red
	dashed lines to show the ideal well-resolved growth rate. For an easier comparison, we have multiplied
	the $x$-position of the low resolution simulation ($\mathcal{R} = 2$) by
	3/11 so that the wavelength of the perturbation matches that of the
	resolved simulation. 
	In the right column we show the resolved simulation but with the full vertical scale for reference. 
	From top to bottom, the times are $\tau = 0, 0.514, 0.886, 1.329, 1.542,$ and $1.772$, respectively. 
	}
	\label{fig:evolution}
\end{figure*}

\RED{To show that the fluctuations in the unresolved simulations indeed do
not oscillate but actually grow as expected, we show the longer timescale evolution in Fig.~\ref{fig:evolution}. 
We show TSPH and MFV simulations with low resolution $\mathcal{R} = 2$ and a 
perturbation wavelength of  $\lambda/\lambda_\mathrm{J}=1.5$. 
We also show a higher resolution simulation (with $\mathcal{R} = 0.25$) with MFV for a comparison of the expected perturbation growth.
We multiply the $x$-positions of 
the low resolution simulations ($\mathcal{R} = 2$) by 3/11 so that the 
wavelength of the perturbation matches that of the resolved simulation, for ease of comparison.
}

\RED{
With SPH, the initial density perturbation slowly grows, keeping its 
sinusoidal shape. 
The growth is, however, much slower than that of the well-resolved simulation (red dashed lines). These perturbations fragment into shorter wavelength perturbations after $\tau \simeq 1.54$. 
}

\RED{
With MFM, as we mentioned before, the initial density perturbation fails to
grow. The newly developed density perturbations keep growing and the peak
density fluctuation is higher than with SPH at $\tau = 1.772$. The peak
density is, of course, much smaller than in the resolved simulation; the
difference is more than an order of magnitude.
}

\RED{ Note that, even in the well-resolved run, the growth is slowed when
density reaches too high to resolve the Jeans length with this resolution 
because, for isothermal gas, the Jeans length, $\lambda_\mathrm{J} \propto \rho^{-1/2}$, is a faster
decreasing function of density than the kernel size, $h \propto \rho^{-1/3}$. 
When the density becomes too high, fragmentation into smaller wavelengths is 
also observed for this simulation, as we would expect from the behaviour of the lower resolution simulations. 
}

\RED{We have confirmed that, in all
these simulations, the density perturbations grow until the timesteps become
too small and the code ultimately crashes.}

\section{Discussion and conclusions}
\label{conclusions}

We have carried out the Jeans test using Lagrangian hydrodynamic schemes
implemented in {\scriptsize GIZMO}: TSPH, PSPH, MFM, and MFV. 
We have confirmed that the
TSPH results are consistent with those of \citet{hubber2006}. 
The PSPH results are indistinguishable from those with
TSPH, reflecting that the Jeans test does not involve any discontinuities or
shear flows. 

Using the cubic spline kernel, the oscillation periods and the growth timescales with
all the methods converged to the analytic estimates at a resolution of
$\mathcal{R} = 0.25$, where $R$ is the ratio of kernel diameter to perturbation wavelength.  
At $\mathcal{R} <1$, the oscillation period becomes
shorter and the growth timescale longer as the resolution is decreased. 
Convergence to the analytic solution is fastest with the SPH methods and slowest 
with MFV. 
\RED{In all the test simulations, MFM shows a behaviour intermediate between SPH methods and MFV.}

\RED{Unlike Eulerian methods such as AMR,} even when we strongly violate the Jeans 
condition (e.g. $\mathcal{R} = 2$),
none of the four schemes show artificial fragmentation for shorter wavelength
perturbations with $\lambda < \lambda_\mathrm{J}$, i.e. perturbations with a
wavelength shorter than the Jeans length never fragment. This result is
consistent with the SPH-specific investigation of \cite{hubber2006}.
\RED{We suspect that one of the reasons why the Lagrangian methods 
behave differently from AMR is that, in 
the Lagrangian methods we have tested here, the spatial and mass resolution 
are directly coupled, whereas, in AMR, the relation between the two depends 
on the refinement criteria.}

\RED{Longer wavelength perturbations ($\lambda > \lambda_\mathrm{J}$), however,
grow quite differently from the converged behaviour with MFM 
when we fail to satisfy the Jeans condition.
Initially imposed perturbations fail to grow and waves are reflected at the
density peaks.
Newly formed density perturbations during the initial density peaks then grow
and collapse. Interestingly, SPH shows more consistent behaviour with the
converged simulations even when the Jeans condition is violated. Initially
imposed perturbations grow, although the growth is much slower than in the 
resolved cases, and the fragmentation into shorter wavelength fluctuations 
occurs slower than in the equivalent MFV simulation. }

\RED{
	Our results suggest that, when unresolved, the ratio of the magnitude of 
	hydrodynamic force to that of self-gravity at sub-resolution scale 
	is the largest in MFV and smallest in SPH. 
	This sub-resolution scale pressure gradient force in MFV overcomes the softened self-gravitational
	force when we impose unresolved strongly nonlinear initial density perturbations. 
	The spectrum between the three Lagrangian methods is consistent with 
	the fundamentals of the schemes. 
	Of the three, MFV is the closest to Eulerian methods in the way that it
	employs a finite volume method with a Riemann solver like AMR,  with
	inter-cell mass fluxes. MFM lies between SPH and MFV: like MFV, 
	MFM uses a finite volume method with a Riemann solver, but it is similar 
	to SPH in that there is no inter-cell mass flux.
}
 
\RED{
The difference between SPH and MFV when we impose strongly non-linear unresolved
perturbations may be consistent with the kernel-scale difference in
dissipation and noise in SPH and MFV demonstrated in \citet[][Section~4.4.4]{hopkins2015}. 
In the power spectrum of driven, isothermal turbulence, they find a `bottleneck' feature
(excess power on scales just above the dissipation range) in MFM and MFV simulations. 
On the other hand, with SPH, the power spectrum falls below the expected value. 
The differences between SPH and MFV are qualitatively similar to the difference we report here.}
\RED{
	In their test, however, MFM and MFV behave indistinguishably. This is in
	contrast to our results, in which MFM is qualitatively more similar to
	SPH than MFV (see Fig.~\ref{fig:final}). The main difference between
	finite mass methods like SPH and MFM and finite volume methods like MFV
	(and AMR) is the existence of the inter-cell mass fluxes. 
	In a method with mass
	fluxes where gravity and hydro are operator split, it is difficult to
	maintain hydrostatic equilibrium \citep[e.g.][]{Zingale_2002}. 
	To realise the hydrostatic equilibrium with such a method, large hydrodynamic 
	fluxes and gravity need to cancel out each other; achieving this is difficult with 
	the finite volume method in general. 
	On the other hand, in SPH and MFM, the gravitational force is calculated in a point-like manner,
	approximating a fluid element as a point mass.
	This means that in a method like SPH where the fluid forces are also
	evaluated in a point-like collocated manner one can get much better
	force cancellation. 
	In MFM the introduction of the Riemann solver and the finite-volume 
	method would lead to some deviations. It, however, still has 
	velocity collocation (i.e. no mass fluxes), and this may help 
	to achieve hydrostatic equilibrium. }

We also find that our results obtained with the cubic spline kernel can be
generalised to other kernels by using the effective smoothing length, $h' = 2
\sigma$, where $\sigma$ is the kernel standard deviation defined by
Eq.~(\ref{eq:sd}). By using this effective smoothing length, we can define
the kernel-independent resolution, $\mathcal{R}'$, as $\mathcal{R}' = 4
h'/\lambda$. To follow the growth of a perturbation whose wavelength is close
to the Jeans length with the cubic spline kernel, the resolution required is
$\mathcal{R} = 0.25$. This corresponds to $\mathcal{R}' \simeq 0.28$ and all
the schemes tested here produce almost identical results at this resolution.

\section*{Acknowledgements}
We are grateful to Phillip Hopkins for making {\scriptsize GIZMO} public. 
We thank an anonymous referee for his/her careful and detailed comments on our manuscript. 
We also thank Alex Pettitt for helpful comments. 
Numerical calculations were performed in part using 
Oakforest-PACS at the CCS, University of Tsukuba and cray XC50 at CfCA, NAOJ, and the computer resource offered
under the category of General Project by Research Institute for
Information Technology, Kyushu University.
TO acknowledges the financial support of MEXT KAKENHI Grant 18H04333, 
19H01931, and 20H05861. 
This work is supported by MEXT as ``Program for Promoting Researches on
the Supercomputer Fugaku'' (Toward a unified view of the universe: from
large scale structures to planets).

\section*{Data Availability}
The data underlying this article will be shared on reasonable request to the corresponding author. 








\bsp	
\label{lastpage}
\end{document}